\documentclass[aps,twocolumn,pra,superscriptaddress,showpacs,tightenlines]{revtex4}

\usepackage{amssymb}
\usepackage{amsmath}
\usepackage{graphicx}
\usepackage{epsfig}
\usepackage{subfigure}
\usepackage{amsfonts}

\begin{document}

\title{Quantum super-cavity with atomic mirrors}
\author{Lan Zhou}
\affiliation{Advanced Science Institute, The Institute of Physical
and Chemical Research (RIKEN), Wako-shi 351-0198, Japan}
\affiliation{Department of Physics, Hunan Normal University,
Changsha 410081, China}
\author{H. Dong}
\affiliation{Institute of Theoretical Physics, The Chinese Academy
of Sciences, Beijing, 100080, China}
\author{Yu-xi Liu}
\affiliation{Advanced Science Institute, The Institute of Physical
and Chemical Research (RIKEN), Wako-shi 351-0198, Japan}
\affiliation{CREST, Japan Science and Technology Agency (JST),
Kawaguchi, Saitama 332-0012, Japan}
\author{C. P. Sun}
\affiliation{Advanced Science Institute, The Institute of Physical
and Chemical Research (RIKEN), Wako-shi 351-0198, Japan}
\affiliation{Institute of Theoretical Physics, The Chinese Academy
of Sciences, Beijing, 100080, China}
\author{Franco Nori}
\affiliation{Advanced Science Institute, The Institute of Physical
and Chemical Research (RIKEN), Wako-shi 351-0198, Japan}
\affiliation{CREST, Japan Science and Technology Agency (JST),
Kawaguchi, Saitama 332-0012, Japan} \affiliation{Center for
Theoretical Physics, Physics Department, Center for the Study of
Complex Systems, The University of Michigan, Ann Arbor, MI
48109-1040, USA.}

\begin{abstract}
We study single-photon transport in an array of coupled
microcavities where two two-level atomic systems are embedded in two
separate cavities of the array. We find that a single-photon can be
totally reflected by a single two-level system. However, two
separate two-level systems can also create, between them,
single-photon quasi-bound states. Therefore, a single two-level
system in the cavity array can act as a mirror while a different
type of cavity can be formed by using two two-level systems, acting
as tunable ``mirrors'', inside two separate cavities in the array.
In analogy with superlattices in solid state, we call this new
``cavity inside a coupled-cavity array'' a super-cavity. This
supercavity is the quantum analog of Fabry-Perot interferometers.
Moreover, we show that the physical properties of this quantum
super-cavity can be adjusted by changing the frequencies of these
two-level systems.
\end{abstract}

 \maketitle \narrowtext

\section{\label{Sec:1}Introduction}

In quantum networks, photons provide faithful quantum information transfer,
because they travel at the speed of light over long distances, and with
little decoherence compared to other information carriers (e.g., electrons).
To interconnect networks, it is crucial to have a quantum memory at the
switching nodes. Many approaches have been proposed to realize quantum
memories, where quantum information can be stored and retrieved, for
instance, using electromagnetically-induced transparency (e.g., in Refs.~
\cite{Harris,Lukin1,scp03}) or photon echoes (e.g., Refs.~\cite%
{phoech1,phoech2,CRIB}). Photons can be confined to a very small
volume (e.g., Ref.~\cite{cavity1}) using micro-cavities or
micro-resonators with low dissipation and thus the micro-cavities
can serve as quantum memories. Moreover, experiments also
demonstrated that the quality factor of a photonic crystal
nanocavity~\cite{cavity3,cavity4} or microwave
cavity~\cite{cavity4a,cavity4b} can be controlled by dynamically
changing the environment of the cavity.

To faithfully transfer quantum information, individual photon control would
be desirable. Single-photon turnstiles have been studied in, e.g., Refs.~%
\cite{cavity,cavity2,cavity5}. There, a semiconductor quantum dot~\cite%
{cavity} or a single atom~\cite{cavity2,cavity5} can behave as a photon
turnstile. Recently, a nonlinear two-photon switch device using nanoscale
surface plasmons has been~\cite{Lukin-np} theoretically studied, where a
single-photon `gate' is used to control the propagation of subsequent single
photons. Considering the one-dimensional scattering process of
single-photons by a two-level system, the total reflection can be controlled
by tuning the inner structure of the scatterer~\cite{fanpaper,ZGLSN,cavity6}%
. A solid state device, functioning as a single-photon quantum switch~\cite%
{Sun1,Mari} in a one-dimensional coupled-cavity waveguide has been
studied, also using a discrete-coordinate approach~\cite{ZGLSN}.

For single-photon transport in a one-dimensional waveguide, the
photons can be totally reflected~\cite{fanpaper,ZGLSN,cavity6} by a
controllable two-level system which can act as a perfect mirror. It
is known that the Fabry-Perot cavity, which consists of two highly
reflecting planar mirrors, is the simplest cavity. It is then
natural to ask the question: ``is it possible to construct \textit{a
quantum resonator, in a one-dimensional waveguide, with two
controllable quantum scatterers?}'' Here, we focus on this question
and study \textit{quantum analogs of the Fabry-Perot cavity}.

This paper studies the coherent transport of photons, which
propagate in a one-dimensional coupled-resonator waveguide (CRW) and
are scattered by two controllable two-level systems located
separately in the CRW. Besides presenting a unified theory,
including both the long-wavelength and short-wavelength regimes, the
discrete coordinate approach employed in this work shows that:
\begin{enumerate}
\item Photon quasi-bound states, with a tunable leakage, appear in the region
sandwiched between the two two-level systems, when the interaction
between the two-level systems and the cavity field is strong
compared with the hopping constant. (Hereafter, for brevity, we will
often use the word ``atom'' instead of ``two-level system''. In the
terminology of quantum information, a two-level system is a qubit,
therefore, hereafter, a two-level system is sometimes called a
qubit, but this ``atom'' refers to an artificial atom made, e.g.,
from a superconducting circuit.)
\item A perfect quantum super-cavity, confining photons inside the two
``atoms'', can be formed when the transition energies of these two-level
systems are equal to the photon energy, with wave number $k=n\pi /(2d)$, where $%
2d$ is the distance between the two atomic scatterers and $n$ is an
integer.
\item Photons can be stored and re-emitted by adjusting the transition
frequencies of these two two-level systems.
\end{enumerate}

This paper is organized as follows: in Sec.~\ref{Sec:2}, we present
our model, a coupled-cavity array with two atoms separately embedded
in two different cavities. In Sec.~\ref{Sec:3}, we study the
transport properties of a single-photon, and derive the conditions
for the coherent control of a single-photon scattering by two atoms.
In Sec.~\ref{Sec:4}, the quantum super-cavity, with \textit{two
atomic mirrors}, is studied. We prove that the leakage of this
super-cavity is tunable by changing the transition energy of these
two atoms. The wave numbers inside this super-cavity are also
analytically obtained by a perturbation approach. Moreover, we study
how a super-cavity can be formed in the long-wavelength (low-energy)
regime, in Sec.~\ref{Sec:5} and in the short-wavelength
(higher-energy) regime, in Sec.~\ref{Sec:6}. These two regimes
correspond to the quadratic and linear photon dispersion relation,
respectively. Conclusions are summarized at the end.

\section{\label{Sec:2}model}

\begin{figure}[tbp]
\includegraphics[bb=73 340 545 547, width=8.5 cm,clip]{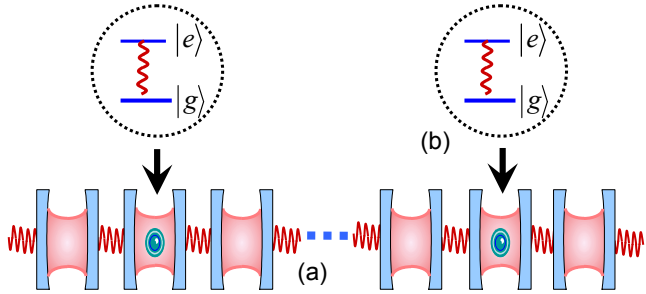}
\caption{(Color online). Schematic configuration for a quantum super-cavity
realized by two atoms embedded in two separated cavities of a coupled
resonator waveguide, as shown in (a). Each atom can be represented as in
(b). These two atoms can behave as two partly-reflecting mirrors, forming a
cavity-within-a-cavity, or super-cavity. This can act as a quantum analog of
Fabry-Perot interferometer.}
\label{fig2:1}
\end{figure}

As shown in Fig.~\ref{fig2:1}, we consider a one-dimensional
coupled-resonator waveguide (CRW) with two two-level systems,
embedded separately in two distant cavities. The CRW can be either
an array of coupled superconducting transmission line resonators or
an array of coupled defect resonators in photonic crystals (see,
e.g. Ref.~\cite{greentree}). The two-level systems can be either
natural atoms or artificial atoms (e.g., superconducting qubits or
semiconducting quantum dots).
In contrast to the similar configurations in
Refs.~\cite{AngelA76,zhjsun76,hu76PRA,Plenio,Sergey}, here only two
atoms are located inside the CRW.

Once a photon is inside one cavity of the CRW, it propagates along the CRW
and is also scattered by the atoms. The CRW can be described by the
Hamiltonian
\begin{equation}
H_{c}=\omega \sum_{j=-\infty}^{\infty}a_{j}^{\dagger }a_{j}-\xi
\sum_{j=-\infty}^{\infty}\left( a_{j}^{\dagger
}a_{j+1}+\mathrm{h.c.}\right) \label{2model-01}
\end{equation}
with the annihilation operator $a_{j}$ and creation operator $a_{j}^{\dag }$
of the $j$th cavity mode. The first term of Eq.~(\ref{2model-01}) denotes
the free Hamiltonian of all the resonators. The second term of Eq.~(\ref%
{2model-01}) represents the couplings between any two nearest-neighbor
cavities. For example, $a_{j}a^{\dagger}_{j+1}$ means that the photon is
annihilated in the $j$th cavity and is created in the $(j+1)$th cavity.
Here, for simplicity, we assume that all resonators have the same frequency $%
\omega $ and the hopping energies $\xi$ between any two nearest-neighbor
cavities are the same. The hopping energy $\xi$ is determined by the
inter-cavity coupling. $H_{c}$ is a typical tight-binding boson model and
can be rewritten as
\begin{equation}
H_{c}=\sum_{k}E_{k} b^{\dagger}_{k}b_{k}
\end{equation}
by introducing the Fourier transform
\begin{equation}
b_{k}=\frac{1}{\sqrt{N}}\sum_{j}e^{ikj}a_{j}.
\end{equation}%
The dispersion relation between $E_{k}$ and $k$ is given by
\begin{equation}
E_{k}=\omega -2\xi \cos k\text{,}  \label{2model-02}
\end{equation}%
which forms an energy band. Here, the lattice constant $l$ is set to unity.

Let us assume that each atom has a ground state $\left\vert
g\right\rangle$ and an excited state $\left\vert e\right\rangle$.
Let the distance between the two atoms embedded in the CRW be $2d$.
For convenience, we take the $0$th cavity as the coordinate-axis
origin. As shown in Fig.~\ref{fig2:1}, we also assume that the first
atom, with transition energy $\Omega_{1}$, is located at the
$(-d)$th cavity, on the left side of the origin, and the second
atom, with transition energy $\Omega _{2}$, is embedded in the $d$th
cavity, on the right side of the origin. Under the rotating-wave
approximation, the interaction between the $d$th and $(-d)$th cavity
fields and the two atoms is described by the Jaynes-Cummings
Hamiltonian
\begin{equation}
H_{I}=\sum_{l=1,2}\left[ \Omega _{l}\left\vert e\right\rangle
_{l}\left\langle e\right\vert +J_{l}\left( \left\vert e\right\rangle
_{l}\left\langle g\right\vert a_{\left( -1\right) ^{l}d}+\mathrm{h.c.}%
\right) \right]  \label{2model-03}
\end{equation}%
where $J_{l}$ is the coupling strength between the $l$th atom and
the $(-1)^l d$th cavity field.

The total Hamiltonian $H=H_{I}+H_{c}$ exihibits different behaviors
in the long-wavelength (low-energy) regime and the short-wavelength
(higher-energy) regime, which correspond to the quadratic and linear
regimes of the photon dispersion relation, respectively. Namely, in
the low-energy regime, the long-wavelength approximation gives a
photon quadratic spectrum
\begin{equation}
E_{k}\simeq \omega -2\xi +\xi k^{2},
\end{equation}
while in the higher-energy regime, the short-wavelength
approximation leads to a photon linear spectrum
\begin{equation}  \label{eq:4}
E_{k}\simeq \omega -\pi \xi \pm 2\xi k.
\end{equation}
Both regimes will be studied in this paper.

We note that an ideal system without losses is considered here. In
practice, both photons and atoms unavoidably interact with different
environments, that is, dissipation always exists. The dissipation
substantially reduces the propagating length of the photons, and so
does the transmission of the single-photon. In order to present the
main physics of this system, we neglect dissipation, decoherence,
and the nonuniform couplings in this paper. These effects are
separately studied and will be presented in the future.

\section{\label{Sec:3}Single-Photon Reflection and transmission}

A photon incident from the left of the CRW, with energy within the energy
band, propagates along the one-dimensional CRW until it is scattered by the
first atom. Then it splits into a transmitted and a reflected portions. The
transmitted part propagates freely until it encounters the second atom,
where the same type of splitting occurs once again.

In this section, we will discuss the reflection and transmission
coefficients of a single-photon in terms of the projection of the
asymptotic wavepackets onto appropriate plane waves. First, we
consider the eigenstates of the total system. Three
mutually-exclusive possibilities are considered: either the photon
is propagating inside the cavity, or the photon is absorbed by one
atom or the other. Considering all of three cases, the stationary
state for the Hamiltonian $H=H_{I}+H_{c} $ is written in the form
\begin{equation}
\left\vert E_{k}\right\rangle =\sum_{j}u_{k}\left( j\right)
a_{j}^{\dag }\left\vert 0gg\right\rangle +u_{1e}^{k}\left\vert
0eg\right\rangle +u_{2e}^{k}\left\vert 0ge\right\rangle,
\label{2RTA-01}
\end{equation}%
where the first number $0$ inside the Dirac brakets represents the
vacuum state of all cavity fields. The parameter $u_{k}\left(
j\right) $ represents the probability amplitude for finding the
photon at the $j$th cavity. $u_{le}^{k}$ is the probability
amplitude of the $l$th atom in its excited state while the other
atom is in the ground state and all the cavity fields are in the
vacuum. This form of $|E_{k}\rangle$ includes the three cases listed
above. Using the Schr\"{o}dinger equation, the single-photon
scattering process can be described by the following equation
\begin{eqnarray}
&&\left( E_{k}-\omega \right) u_{k}\left( j\right) +\xi \left[ u_{k}\left(
j+1\right) +u_{k}\left( j-1\right) \right]  \label{2RTA-02} \\
&&=J_{1}G_{1k}\delta _{j(-d)}u_{k}\left( -d\right)
+J_{2}G_{2k}\delta _{jd}u_{k}\left( d\right),  \notag
\end{eqnarray}%
where the Green function $G_{lk}=J_{l}/\left( E_{k}-\Omega _{l}\right) $ is
obtained from the relation
\begin{equation}
u_{le}^{k}=\frac{J_{l}}{E_{k}-\Omega _{l}}u_{k}\left[ \left( -1\right) ^{l}d%
\right] \text{.}  \label{2RTA-03}
\end{equation}%
If we regard the second term of the left side of Eq.~(\ref{2RTA-02}) as the
kinetic energy term and the right side of Eq.~(\ref{2RTA-02}) as potential
energy term, then Eq.~(\ref{2RTA-02}) describes the eigenfunction $%
u_{k}\left(j\right)$ subjected to a potential with singularities at $j=\pm d$%
. In the region $j\neq \pm d$, the potential is zero, the solutions to Eq.~(%
\ref{2RTA-02}) are plane waves with wave-vectors $k$. Therefore we consider
the wave functions
\begin{equation}
u_{k}\left( j\right) =\left\{
\begin{array}{c}
e^{ikj}+re^{-ikj}\text{, \ \ \ \ \ }j<-d, \\
Ae^{ikj}+Be^{-ikj}\text{, }-d<j<d, \\
te^{ikj}\text{, \ \ \ \ \ \ \ \ \ \ \ \ \ }j>d.%
\end{array}%
\right.  \label{2RTA-04}
\end{equation}

The $u_{k}\left(j\right)$ in Eq.~(\ref{2RTA-04}) describes the
scattering process of an initial plane wave $\exp \left( ikj\right)$
incident from the left side of $j=-d$. This freely propagating wave
is either reflected or transmitted when it encounters the first
scatterer. The reflection and transmission are described by the
amplitudes $r$ and $A$. The transmitted wave propagates freely until
it encounters the second scatterer at the point $j=d$, where the
corresponding reflection and transmission amplitudes are described
by $B$ and $t$. Here both scatterers produce a highly localized
repulsive or attractive effective force, which depends on the
incident energy of the single photon, the transition energies of the
atoms, and the coupling strength between the atoms and their
corresponding cavities.

\subsection{atomic transition energy $\Omega_{l}$ inside the band}

\begin{figure}[tbp]
\includegraphics[bb=70 437 558 593, width=7 cm,clip]{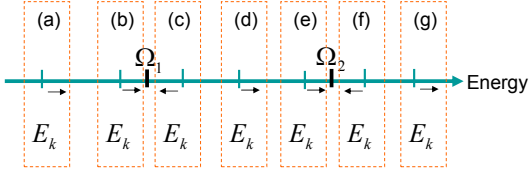}
\caption{(Color online). This figure schematically illustrates seven
different cases (a-g) discussed in the text and also corresponding
to the seven cases (a-g) in Fig.~\protect\ref{fig2:2}. Here $E_{k}$
is the photon energy, and $\Omega_{1},\Omega_{2}$ are the two atomic
transition energies. The arrows indicate when the photon energy
$E_{k}$ increases or decreases. In: (a) $E_{k}<\Omega_{1}$; (b)
$E_{k}$ is slightly below $\Omega_{1}$, and increasing; (c) $E_{k}$
is slightly above $\Omega_{1}$, and decreasing; (d) $
\Omega_{1}<E_{k}<\Omega_{2}$; (e) $E_{k}\rightarrow\Omega_{2}$; (f)
$E_{k}$ approaches $\Omega_{2}$ from the right side, (g) $E_{k}$
moving away from $\Omega_{2}$.} \label{fig2:2a}
\end{figure}

\begin{figure}[tbp]
\includegraphics[bb=97 163 457 706, width=8 cm,clip]{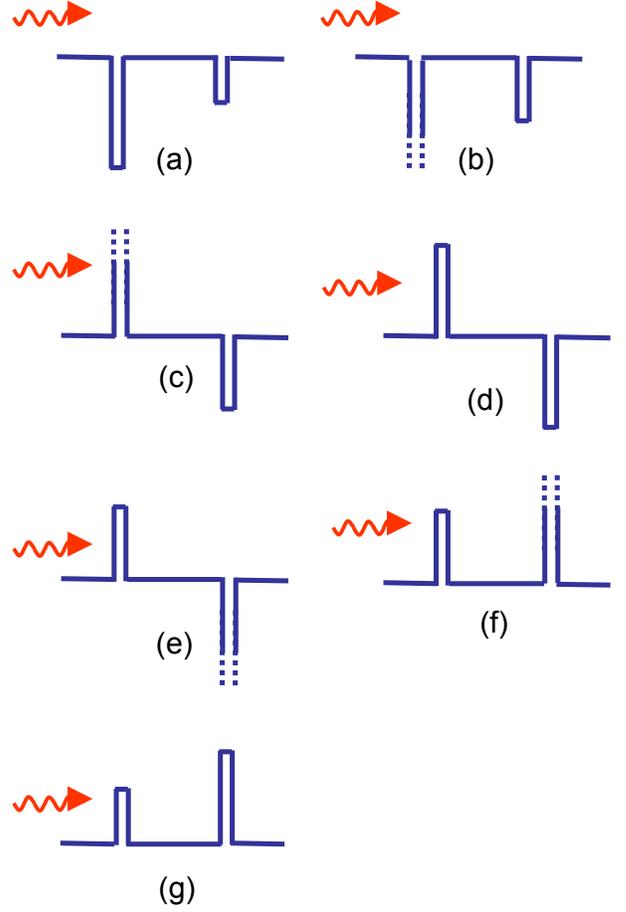}
\caption{(Color online). Schematic diagrams of the single-photon
scattering process with $\Omega _{1}<\Omega _{2}$ inside the energy
band. Here, the vertical axis is energy and the horizontal axis is
the position along the CRW. Also, $\Omega_{l}$ is the transition
energy of the $l$-th atom ($i=1,2$). The dashed lines refer to an
infinitely high potential barrier or well. The seven cases (a-g)
shown here correspond to the seven photon energy regimes (a-g)
explained in Fig.~\protect\ref{fig2:2a}. The cases (b), (c), (e),
and (f) correspond to the total reflection of the incident photon
because in each one of these cases one of the wells or barriers is
infinite.} \label{fig2:2}
\end{figure}

Figure~\ref{fig2:2a} schematically illustrates the seven cases (a-g)
shown in Fig.~\ref{fig2:2}. Figure~\ref{fig2:2} shows how the
potential energy depends on the energy $E_{k}$ of the incident
photon. All the results schematically shown in Fig.~\ref{fig2:2} are
derived directly from Eq.~(\ref{2RTA-02}). Here, we assume $\Omega
_{1}<\Omega _{2}$ and both transition energies $\Omega_{l} (l=1,2)$
are in the region $[\omega -2\xi ,\omega +2\xi ]$. When the photon
energy $E_{k}<\Omega _{1}$ is smaller than the transition energy of
the first atom, two attractive delta function potentials appear at
$j=\pm d$ (as schematically shown in Figs.~\ref{fig2:2a}a
and~\ref{fig2:2}a). If the photon energy $E_{k}$ is increased and
approaches $\Omega _{1}$, Figs.~\ref{fig2:2a}(b)
and~\ref{fig2:2}(b), the potential located at $j=-d$ tends to minus
infinity. However, when $E_{k}$ approaches $\Omega _{1}$ from the
right side, as shown in Figs.~\ref{fig2:2a}(c) and ~\ref{fig2:2}(c),
the first potential gradually becomes an infinite barrier. When the
incident photon energy $E_{k}$ is further increased and is between
$\Omega _{1}$ and $\Omega _{2}$, Fig.~\ref{fig2:2a}(d), photons
first collide with a repulsive finite potential, then go through an
attractive potential well, as schematically shown in
Fig.~\ref{fig2:2}(d). As $E_{k}$ further increases,
Fig.~\ref{fig2:2a}(e), the height of the potential barrier at $j=-d$
becomes lower and lower, and the second two-level system creates a
potential well with its depth becoming deeper and deeper,
Fig.~\ref{fig2:2} (e), eventually becoming a potential well with
infinite depth. After $E_{k}$ goes across the transition energy
$\Omega_{2}$, Fig.~\ref{fig2:2a}(f), a double-barrier is produced by
the atoms, as shown in Fig.~\ref{fig2:2a}(f). In this case, the
waves are totally reflected when the height of the delta potential
of the second atom, goes to infinite, Fig.~\ref{fig2:2}(f).
Figure~\ref{fig2:2}(g) shows the potential energy corresponding to
the energies shown in Fig.~\ref{fig2:2a}(g). Therefore,
Figure~\ref{fig2:2} schematically presents ways to control photon
transport by, e.g., adjusting the transition frequencies,
$\Omega_{1}$ and $\Omega_{2}$, of the two atoms.

From the continuity conditions $u_{k}\left( \pm d^{+}\right)
=u_{k}\left( \pm d^{-}\right) $ and the eigenvalue equations
\begin{eqnarray}
&&\left( E_{k}-\omega -J_{2}G_{k2}\right) u_{k}\left( d\right)  \label{2RTA-05a} \\
&&=-\xi \left[ u_{k}\left( d+1\right) +u_{k}\left( d-1\right) \right] , \notag\\
\notag \\
&&\left( E_{k}-\omega -J_{1}G_{k1}\right) u_{k}\left( -d\right)  \label{2RTA-05b} \\
&&=-\xi \left[ u_{k}\left( -d+1\right) +u_{k}\left( -d-1\right)
\right]\notag
\end{eqnarray}%
at $j=\pm d$, the transmission amplitude $t$ can be derived:
\begin{eqnarray}
t &=&4\xi ^{2}\sin ^{2}k [ \left( e^{i4kd}-1\right) J_{1}G_{k1}J_{2}G_{k2}
\label{2RTA-06} \\
&&+2i\xi \sin k\sum_{l}J_{l}G_{kl}+4\xi ^{2}\sin ^{2}k] ^{-1}\text{.}  \notag
\end{eqnarray}
Above, $d^{\pm}=d\pm \epsilon$, where $\epsilon$ is a very small
positive number. When the photon frequency $E_{k}$ matches one of
the atomic transition frequencies $\Omega _{l}$ ($l=1,2$), the
transmission $t$ is zero. This $t=0$ case occurs in cases (b), (c),
(e), and (f) in Figs.~\ref{fig2:2a} and~\ref{fig2:2}. When an atom
has its transition energy $\Omega_{l}$ inside the energy band, it
may be excited by the incident photon. The absorption or emission of
a photon by an atom leads to wave interference between the incident
wave and the reflected wave.

\subsection{atomic transition energy $\Omega_{l}$ outside the band: $%
\Omega_{l}<\omega-2\xi, \Omega_{l}>\omega+2\xi$}

For an atom with transition energy $\Omega_{l}$ far away from the
energy band, the propagating single photon cannot excite the atom,
thus the photon emerges in the other side of the scatterers with its
energy (almost) equal to its original one, due to energy
conservation.

\begin{figure}[tbp]
\includegraphics[width=8 cm]{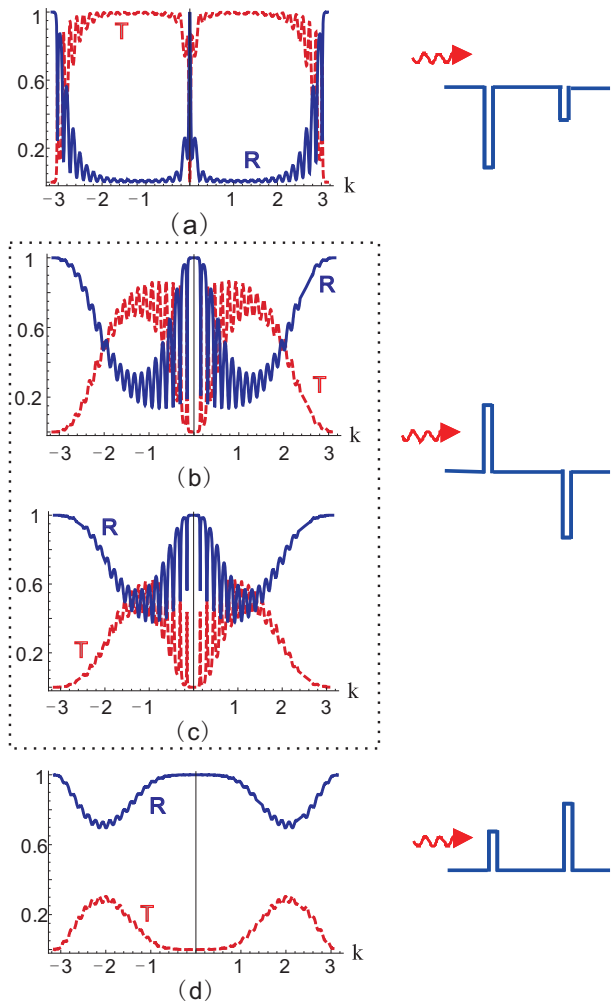}
\caption{(Color online). The photon reflection coefficient $R=1-T$
(blue solid line) and the photon transmission coefficient, $T$ (red
dashed line) as a function of the photon wave number $k$ ($-\pi\leq
k\leq \pi$) when both atomic transition frequencies $\Omega_{l}$ are
outside the band $[\omega-2\xi,\omega+2\xi]$. Here, $k$ and $d=10$
is in units of the lattice constant, and other parameters are in
units of $\xi$. Here, the cavity energy is $\omega=5$, (a) $\Omega
_{1}=8$, $\Omega _{2}=8$, and the coupling between the atoms and the
CRW are $J_{1}=0.5$, $J_{2}=0.7$; (b) $\Omega _{1}=2$, $\Omega
_{2}=8$, $J_{1}=0.7$, $J_{2}=2$; (c) $\Omega _{1}=2$, $\Omega
_{2}=8$, $J_{1}=0.7$, $J_{2}=2.6$; (d) $\Omega _{1}=2$, $\Omega
_{2}=2.7$, $J_{1}=0.5$, $J_{2}=3$. Panel (a) corresponds to two
potential wells at $j=\pm d$. Panels (b) and (c) correspond to one
potential barrier at $j=-d$ and one potential well at $j=d$. Panel
(d) corresponds to two potential barriers.} \label{fig2:3}
\end{figure}

Using Eq.~(\ref{2RTA-06}), in Fig.~\ref{fig2:3}, we plot the
reflection coefficient $R=1-T$ (blue solid line) and the
transmission coefficient $T=|t|^2$ (red dashed line) as a function
of the incident photon wave number $k$ ($-\pi\leq k \leq \pi$), when
both atomic transition energies $\Omega_{l}$ are outside the energy
band. As shown in Fig.~\ref{fig2:3}, total reflection $R=1$ always
happens at $k=0,\pm \pi$ for nonzero $J_{1}$ and $J_{2}$, and this
total reflection is completely independent of the transition
energies $\Omega_{l}$ of the atoms. This observation is caused by
the following reasons. First, a band, Eq.~(\ref{2model-02}), is
formed in this periodic CRW, which acts like a photon filter or a
photon ``band-pass filter'': transmitting photons over a limited
frequency range. Incident photons with energy $E_{k}$ outside the
band do not interact with the CRW, therefore the photon group
velocities vanish at the zone boundaries or band edges. Second, when
the atomic transition energies $\Omega_{l}$ are outside the band,
the infinite delta potential wells or barriers cannot be formed, and
the total photon reflection does not occur, except when $k=0,\pm \pi
$.

The oscillations shown in $T(k)$ and $R(k)$ in Fig.~\ref{fig2:3}
originate from the multiple interference of waves in the region
sandwiched by the two atoms. Comparing Figs.~\ref{fig2:3}(a)
and~\ref{fig2:3}(b), we find that as the coupling strengths $J_{l}$
between the atoms and the CRW increase ($J_{1}$ from $0.5$ to $0.7$
and $J_{2}$ from $0.7$ to $2$), the oscillation in $R(k)$ becomes
much larger when $|k|\leq 1$, i.e., increasing the coupling
strengths $J_{l}$ magnifies the oscillations in $R(k)$ and $T(k)$.
Indeed the wave interference giving rise to the oscillations in
$R(k)$ and $T(k)$ varies with five parameters: the energy $E_{k}$ of
the incident photon, the atomic energies $\Omega_{l}$, and the
couplings $J_{l}$ between the atoms and the CRW. The case shown in
Fig.~\ref{fig2:3}(a) corresponds to two potential energy wells. The
cases considered in Figs.~\ref{fig2:3}(b) and~\ref{fig2:3}(c)
correspond to one barrier in $j=-d$ and one well in $j=d$. The case
considered in Fig.~~\ref{fig2:3}(d), there is a double-barrier. As
is well known, the reflection and transmission coefficients are the
same for a delta potential barrier and a delta potential well, but
their reflection and transmission amplitudes are different by a
phase factor. It is this phase difference that produces the clearly
visible oscillations, due to interference shown in
Fig.~\ref{fig2:3}.

Figure~\ref{fig2:3} shows the complex dependence of $R(k)$ and
$T(k)$ as a function of the coupling strengths $J_{l}$, the hopping
energy (or the inter-cavity coupling strength) $\xi$, and the
detunings, $\delta _{l}=\omega -\Omega _{l}$, between the atoms and
their corresponding cavities. If both coupling strengths $J_{l}$ are
much smaller than the hopping energy $\xi$, the hopping plays a
leading role. In this case, and as shown in Fig.~\ref{fig2:3}(a),
the transmission $T(k)$ is quite large. When either $J_{l}$ is
larger than its corresponding detuning $\delta_{l}$, and also larger
than $\xi$, the reflection dominates, as in Fig.~\ref{fig2:3}(d).
This result can also be found from the phase diagram (Fig.~4) in
Ref.~\cite{ZGLSN}. Figures~\ref{fig2:3}(b) and~\ref{fig2:3}(c) show
the intermediate stage between nearly total transmission in
Fig.~\ref{fig2:3}(a) and nearly total reflection in
Fig.~\ref{fig2:3}(d). In order to make this point somewhat explicit,
we now approximately write the transmission amplitude as
\begin{eqnarray}
t &\approx&\sin ^{2}k\left[ \left( e^{i4kd}-1\right)
\frac{J_{1}G_{1}}{2\xi }\frac{
J_{2}G_{2}}{2\xi }\right.  \label{2RTA-06a} \\
&&\left. +i\sin k\sum_{l}\frac{J_{l}G_{l}}{2\xi }+\sin ^{2}k\right]
^{-1} \notag \\
&=& \sin ^{2}k\left[ \left( e^{i4kd}-1\right) \frac{J_{1}^{2}}{2\xi
\delta_{1}}\frac{ J_{2}^{2}}{2\xi \delta_{2} }\right. \notag  \\
&& \left. +i\sin k\sum_{l}\frac{J_{l}^{2}}{2\xi \delta_{l}}+\sin
^{2}k\right] ^{-1} \notag\text{,}
\end{eqnarray}
when the coupling strength $J_{l}$ is larger than the hopping energy
and smaller than the corresponding detuning $\delta_{l}$. Here,
$G_{l}=J_{l}/\delta _{l}$. Equation~(\ref{2RTA-06a}) shows that,
when $J_{l}^{2}\gg 2\xi \delta _{l}$, the reflection spectrum is
much larger than the transmission spectrum, which coincides with the
change shown numerically from Fig.~\ref{fig2:3}(b) to
Fig.~\ref{fig2:3}(c). Notice that in the large detuning condition
($\delta_{l}\gg J_{l}$), the magnitude $J_{l}^{2}/\delta _{l}$ is
the shift of the energy levels due to the atom-cavity interaction
(see Figs.~\ref{fig2:4} and~\ref{fig2:5}). Therefore, the relation
between $J_{l}^{2}/\delta _{l}$ and the half-width $2\xi$ of the
band determines whether the reflection or the transmission plays a
dominate role. This phenomenon will become much clearer from the
energy-level diagrams shown in the next section.

When the atoms embedded in $\pm d$ cavities are identical, e.g.,
$\Omega =\Omega _{1}=\Omega _{2},\, J=J_{1}=J_{2}$, the transmission
coefficient $T=\left\vert t\right\vert ^{2}$ is derived from
Eq.~(\ref{2RTA-06}) as
\begin{equation}
T=\left\{ 1+\left( JG_{k}\right) ^{2}\left[ \frac{JG_{k}\sin \left(
2kd\right) }{2\xi ^{2}\sin ^{2}k}+\frac{\cos \left( 2kd\right) }{\xi
\sin k}\right] ^{2}\right\} ^{-1}\text{,}  \label{2RTA-07}
\end{equation}%
where
\begin{equation}
G_{k}=\frac{J}{E_{k}-\Omega}.  \label{2RTA-07a}
\end{equation}
We note that the maximum magnitude of the transmission coefficient $T$ in
Eq.~(\ref{2RTA-07}) can be achieved when one of the below conditions is
satisfied: (1) the coupling strength $J$ is much smaller than the detuning $%
\delta_{ph}=E_{k}-\Omega$ between the energy $E_{k}$ of the incident
photon and the transition energy $\Omega_{l}$ of each atom for a
given nonzero $\xi$; (2) the coupling strength $J$ is much smaller
than the hopping energy between adjacent cavities for a definite
nonzero detuning $\delta _{ph} $; (3) the condition
\begin{equation*}
\tan \left( 2kd\right) =-\frac{2\xi \sin k}{JG_{k}}
\end{equation*}
is satisfied. This third leads to a resonant tunneling effect, which
will be studied below.

\section{\label{Sec:4} Super-cavity on resonant states}

The above results show that two atoms may act as a potential
double-barrier. Any potential double-barrier can produce a
wavefunction localized in space (see e.g.,~\cite{FNRMP,fcitaly}).
Therefore, photons located in the range $\left[-d,\,d\right]$,
between the two barriers, may bounce back and forth. Thus, this
double-barrier forms a resonator~\cite{FNRMP}. Photons can leak out
of the resonator owing to the finite width and height of the
potential barriers. Therefore the localized state formed by this
potential energy double-barrier is called a quasi-bound state or a
resonant state~\cite{quaspr,arkaB72,scat-pole}. A particle tunnels,
through two energy barriers, when its energy matches (resonates
with) the localized energy level.

In the system we consider here, the photon propagating in this
system encounters a double potential well/barrier, separated by a
distance $2d$. Such a potential can exhibit resonances. In this
section, we derive the conditions for the photons to be trapped
inside the resonator formed by the two atoms. Therefore, a quantum
super-cavity can be formed by two atoms embedded in two separated
cavities of the CRW.

\subsection{photon wave function in the super-cavity}

For simplicity, we now assume that the atoms at $j=\pm d$ are
identical, e.g. $J_{1}=J_{2}=J$ and $\Omega _{1}=\Omega
_{2}=\Omega$. The eigenfunctions in this system satisfy the discrete
scattering equation derived by Eqs.~(\ref{2RTA-05a})
and~(\ref{2RTA-05b})
\begin{eqnarray}
\left( E_{k}-\omega \right) u_{k}\left( j\right) &=&-\xi \left[ u_{k}\left(
j+1\right) +u_{k}\left( j-1\right) \right]  \label{rs-01} \\
&&+JG_{k}\left[ \delta _{j-d}u_{k}\left( j\right) +\delta _{jd}u_{k}\left(
j\right) \right] \text{.}  \notag
\end{eqnarray}%
A resonant state is an eigenfunction of Eq.~(\ref{rs-01}) under the boundary
condition that only outgoing waves appear outside the potential. Therefore
we assume that Eq.~(\ref{rs-01}) has the following solutions
\begin{equation}
u_{k}\left( j\right) =\left\{
\begin{array}{c}
Ce^{-ikj}\text{, \ \ \ \ \ \ \ \ \ \ \ \ }j<-d, \\
A_{b}e^{ikj}+B_{b}e^{-ikj}\text{, }-d<j<d, \\
De^{ikj}\text{, \ \ \ \ \ \ \ \ \ \ \ \ \ }j>d,%
\end{array}
\right.  \label{rs-02}
\end{equation}
where the coefficients $A_{b}$, $B_{b}$, $C$ and $D$ are the
amplitudes for finding the particle in the state $\exp \left( \pm
ikj\right)$ respectively. We also define the normalized amplitudes
via the ratios
\begin{equation*}
b_{b}\equiv \frac{B_{b}}{A_{b}}\text{, }\,\,\, c\equiv
\frac{C}{A_{b}}\text{, } \,\,\, d\equiv \frac{D}{A_{b}}\text{.}
\end{equation*}
By imposing the continuity equation $u_{k}\left( d^{+}\right)
=u_{k}\left( d^{-}\right) $ and using the Schr\"{o}dinger
Eq.~(\ref{rs-01}) at the point $j=d$, we find
\begin{subequations}
\label{rs-02a}
\begin{align}
b_{b}& =\frac{JG_{k}e^{i2kd}}{2i\xi \sin k-JG_{k}}, \\
d& =\frac{i\xi \sin k}{2i\xi \sin k-JG_{k}}\text{.}
\end{align}%
Using the continuity equation $u_{k}\left( -d^{+}\right) =u_{k}\left(
-d^{-}\right)$ and the equation~(\ref{rs-01}) at the point $j=-d$, we have
\end{subequations}
\begin{subequations}
\begin{eqnarray}  \label{rs-02aa}
b_{b} &=&\left( \frac{2i\xi \sin k}{JG_{k}}-1\right) e^{-2ikd}, \\
c &=&e^{-i2kd}\frac{i\xi \sin k}{JG_{k}}.
\end{eqnarray}%
Obviously, Eq.~(\ref{rs-02a}a) and Eq.~(\ref{rs-02aa}) must be
equal, and the odd and even parities of the quasi-bound states in
Eq.~(\ref{rs-02}) are included in
\end{subequations}
\begin{equation}
e^{i2kd}=\pm \left( \frac{2i\xi }{JG_{k}}\sin k-1\right),  \label{rs-02b}
\end{equation}
where the plus sign gives the even parity and the minus sign has an odd
parity.

\subsection{existence of quasi-bound states}

Let us define a parameter $\lambda =2\xi/J^{2}$. Now we solve the
transcendent equation (\ref{rs-02b}) by a perturbation approach up
to second order in parameter $\lambda $, e.g. $O\left( \lambda^{2}
\right) $. First, let us assume that equation (\ref{rs-02b})
possesses a real solution for $k$ only up to first order in $\lambda
$ (This assumption will be proved later in this section). We set
$k=k_{\mathrm{re}}$ for a real wave number. When $k$ is a real
number
\begin{subequations}
\label{rs-02c}
\begin{eqnarray}
\cos \left( 2k_{\mathrm{re}}d\right) &=&\mp\; 1\text{,} \\
\sin \left( 2k_{\mathrm{re}}d\right) &=&\pm\; \frac{2\xi \sin k_{\mathrm{re}}}{%
JG_{k_{\mathrm{re}}}}\text{.}
\end{eqnarray}%
Then the relation
\end{subequations}
\begin{equation}
\tan \left( 2k_{\text{re}}d\right) =-\; \frac{2\xi \sin
k_{\text{re}}}{JG_{k_{\mathrm{re}}}}  \label{rs-03}
\end{equation}%
provides the condition for the existence of quasi-bound levels, which lead
to the transmission coefficient $T=1$. For a resonant state with an odd
parity, the momentum $k$ satisfies
\begin{equation}
k_{\mathrm{re}}d=n\pi -\varepsilon  \label{rs-04}
\end{equation}%
due to the zero probability for finding the particle outside the barriers,
where $\varepsilon $ is a small positive quantity and $n$ is an integer.
Substituting Eq.~(\ref{rs-04}) into the right side of Eq.~(\ref{rs-03}) and
with the condition $J^{2}\gg 2\xi $, the momentum of the resonant state can
be approximately obtained as
\begin{equation}
k_{\mathrm{re}}= q_{n}-\frac{\lambda }{2d}\left( \delta -2\xi \cos
q_{n}\right) \sin q_{n}+\mathcal{O}\left( \lambda^{2}
\right)\text{,} \label{rs-05}
\end{equation}%
where $\delta =\omega -\Omega$, and $q_{n}=n\pi /d$. For even-parity
states, the momenta of the resonant states are similar to that in
Eq.~(\ref{rs-05}) with $n$ replaced by $(n+1/2)$.
Equation~(\ref{rs-05}) implies that discrete levels appear in the
energy band. In a similar way, we can derive the discrete energy of
the resonant states with even parity.

Although Eq.~(\ref{rs-05}) gives the energy of a quasi-bound state,
it fails to describe the behavior of the wavefunction outside the
sandwiched region. Indeed, a complex wave number $k$ must be
considered in order to obtain the lifetime of a quasi-bound state.
To do this, let us come back to Eq.~(\ref{rs-02b}). Here we show
that, to second order in $\lambda$, the imaginary part of $k$ (which
can represent the lifetime of the resonant state via its dispersion
relation), appears when the wave number in Eq.~(\ref{rs-05}) is
treated as a complex number. We obtain approximate analytical
expression of the wave number
\begin{eqnarray}
k &= &q_{n}-\frac{1}{2}Q_{n}+dQ_{n}^{2}  \label{rs-05b} \\
&+&i\frac{\lambda }{2d}Q_{n}\left[ \delta \cos q_{n}-2\xi \cos
\left( 2q_{n}\right) \right]+\mathcal{O}\left( \lambda^{3} \right),
\notag
\end{eqnarray}
where
\begin{equation}
Q_{n}=\frac{\lambda}{d} \left( \delta -2\xi \cos q_{n}\right) \sin
q_{n}\label{rs-05d}.
\end{equation}
\begin{figure}[tbp]
\includegraphics[width=5 cm]{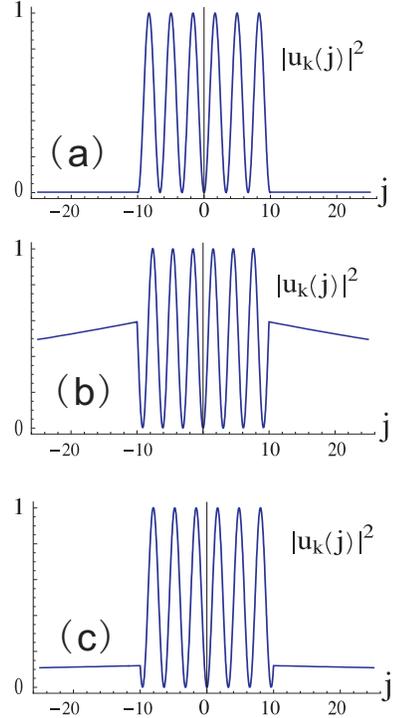}
\caption{(Color online). The probability $|u_{k}(j)|^{2}$ for
finding the photon for a given $k$ versus the position along the
CRW. The parameters are in units of $J$ and are set as follows:
$d=10, \omega=10,n=3, \protect\xi =0.2$, (a) $\Omega =10$, (b)
$\Omega =6$, (c) $\Omega =7 $. $j$ is in units of the lattice
constant. For a given $n$, $q_{n}=n\pi/d$. These were the inputs to
equations~(\ref{rs-05b}) and~(\ref{rs-05d}), which provide
$Q_{n}^{L}$ and $k$. With this $k$, Eqs.~(\ref{rs-02}--\ref{rs-02b})
are used to obtain the $u_{k}(j)$'s shown in the Figures.}
\label{fig2:4}
\end{figure}
Thus, the lifetime of the quasi-bound state with wave number $k$ is
given by the imaginary part of Eq.~(\ref{rs-05b}).

\subsection{quantum super-cavity}

\begin{figure}[tbp]
\includegraphics[bb=133 271 459 628, width=7 cm,clip]{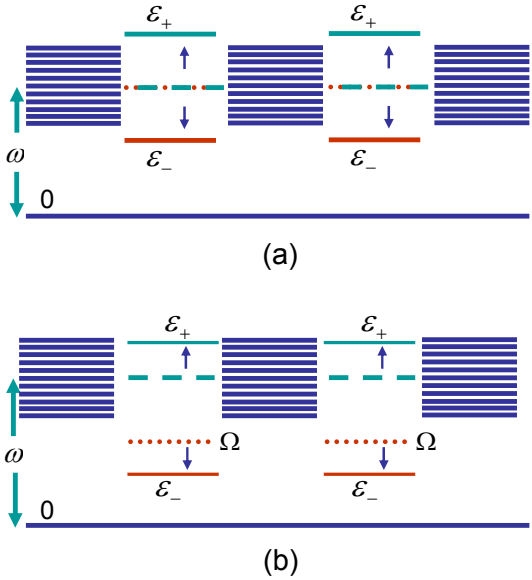}
\caption{(Color online). The corresponding schematic explanation for
the establishment of the super-cavity in Fig.~(\ref{fig2:4}). The
eigenenergy $\omega$ of the cavities at $j=\pm d$ are shown by the
green dashed lines, and the atomic transition energy $\Omega$ is
shown by the red dotted line. The eigenvalues of the two dressed
states are denoted by the symbols $\varepsilon_{\pm}$. The blue
solid lines show the energy band formed by the numerous other
resonators. The arrows denote the shifting of the bare eigenvalues
towards their dressed values $\varepsilon_{\pm}$. Here, (a)
$\omega=\Omega$, (b) $\omega>\Omega$. } \label{fig2:5}
\end{figure}
In Fig.~\ref{fig2:4}, we show the numerically-obtained spatial
distribution of the photon wave function along the CRW for a given
$k$ in Eq.~(\ref{rs-05b}). In Fig.~\ref{fig2:5}, we also give the
corresponding schematic explanation for the establishment of the
super-cavity of Fig.~\ref{fig2:4}. In Fig.~\ref{fig2:5}, the
eigenenergy $\omega$ of the cavity at the site $j=\pm d$ is shown by
green dashed lines, and the atomic transition energy $\Omega$ is
shown by the red dashed lines. The eigenvalues of two dressed states
are denoted by the symbols $\varepsilon_{\pm}$. The blue solid lines
present the energy band formed by other resonators. From
Fig.~\ref{fig2:4}, it can be found that a well-localized state
appears in the sandwiched segment as long as the coupling strength
$J$ is much larger than the hopping energy $\xi$ and the detuning
$\delta =\omega -\Omega$, as shown in Fig.~\ref{fig2:4}(a). In this
situation, the coupling strength $J$ plays a dominant role, and thus
the coupling $J$ shifts the energies at $j=\pm d$ to
\begin{equation}
\varepsilon _{\pm }=\frac{1}{2}\left[ \omega +\Omega \pm
\sqrt{\left( \omega -\Omega \right) ^{2}+4J^{2}}\right] \text{.}
\label{rs-05c}
\end{equation}
When $\omega =\Omega$, the two strong $J$-couplings split the
original degenerate energies of the cavity and the single
atomic-excited state into two new dressed states at the point $j=\pm
d$. These two dressed states are outside the energy band of the
incident photon, as shown in Fig.~\ref{fig2:5}(a). Therefore, the
photons will have a very low probability of going through the atoms,
since the resonance condition is not satisfied in
Fig.~\ref{fig2:5}(a). Thus, if a photon is initially located between
the two atoms, it will remain there, bouncing back and forth from
the atoms. The wavefunctions shown in Figs.~\ref{fig2:4}(b,c) also
indicate that a super-cavity can be formed, but the leakage of this
quantum super-cavity is larger than in Fig.~\ref{fig2:4}(a). The
reason for this large leakage in Fig.~\ref{fig2:4}(c) and
specially~\ref{fig2:4}(b) lies in the energy diagram of
Fig.~\ref{fig2:5}. Although the coupling strength is much larger
than the hopping constant $\xi$ in Figs.~\ref{fig2:4}(b,c), the
original eigen-frequencies, described by $\omega$ and $\Omega$ in
Fig.~\ref{fig2:5}(b), are shifted in opposite directions to
$\varepsilon _{\pm}$, but this shift amount $(\varepsilon
_{+}-\omega)$ is still inside the band, therefore the tunneling
process may appear with larger probability than the case in
Fig.~\ref{fig2:5}(a). Comparing Fig.~\ref{fig2:4}(b) with
Fig.~\ref{fig2:4}(c), it shows that the probability for a
single-photon in the outside region in Fig.~\ref{fig2:4}(b) is
larger than that in Fig.~\ref{fig2:4}(c). Figures~\ref{fig2:4}(b,c)
further show the relation between the magnitude $J^{2}/\delta$ and
the half-width $2\xi$. When $\delta \gg J$ and $J>\xi $, the
dominant photon-atom couplings approximately shifts the energy of
the cavity to
\begin{equation}
\varepsilon_{+}\approx\omega +\frac{J^{2}}{2\delta}.
\end{equation}
Indeed, figures~\ref{fig2:4}(b,c) show the change of the resonant
states when $J^{2}/\delta$ approaches $2\xi$, e.g., the relation
between the dressed state $\varepsilon_{+}$ and the upper edge of
the band. Obviously, the dressed energy level $\varepsilon_{+}$ is
closer to the edge of the band in Fig.~\ref{fig2:4}(c) than the one
in Fig.~\ref{fig2:4}(b), therefore, the probability is much smaller
in Fig.~\ref{fig2:4}(c) for a photon to be outside the sandwiched
segment. Therefore, each atom plays the role of a
partially-reflecting mirror. Here, we present a way to tune the
leakage of the quantum super-cavity.

\subsection{quantum super-cavity with $\Omega$ inside the band.}

The super-cavity was studied above for large coupling strength $J$.
As long as $J$ is nonzero, \textit{a perfect reflection appears when
the transition energy is inside the band}, namely, \textit{when the
single-photon resonates with an atom}~\cite{ZGLSN}. Therefore,
\textit{a perfect super-cavity} exists regardless of the magnitude
of $J$. Of course, a perfect supercavity ($r=1$) is an ideal
limiting case. In reality, decoherence and losses will make the
reflection coefficient $r<1$.

The photon trapping energy can be found analytically, since
Eq.~(\ref{rs-03}) holds exactly when the transition frequency
$\Omega$ satisfies the condition
\begin{equation}
\Omega =\omega -2\xi \cos (q_{n}/2)  \label{rs-06}
\end{equation}%
and the corresponding resonant state has wave numbers $k=q_{n}/2$.
Thus, in this case, the two atoms form two mirrors with perfect
reflection, which leads to a perfect super-cavity.
\begin{figure}[tbp]
\includegraphics[bb=41 300 552 538, width=8 cm,clip]{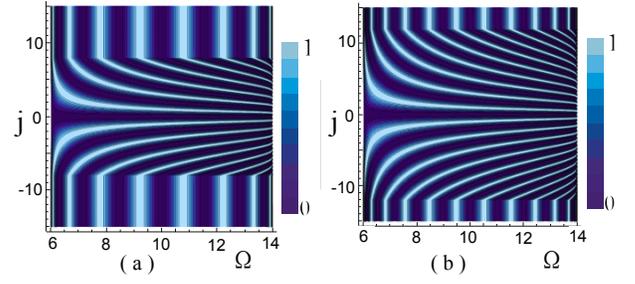}
\caption{(Color online). Contour plots of the norm square of the
wave function $u_{k}(j)$ with odd parity, versus the position along
the CRW and the atomic transition energies $\Omega$, where $\Omega$
is assumed to be inside the band. Here, $\xi =2$, $\omega =10$,
$d=8$ for (a) and $d=12$ for (b).} \label{fig2:6}
\end{figure}
Figure~\ref{fig2:6} shows the contour plots of the probability for a
single-photon as a function of the coordinate $j$ along the CRW and
for the transition energy $\Omega$. Here, the atomic transition
energy $\Omega$ is assumed to be inside the energy band of the CRW.
Also only the wavefunction $u_{k}(j)$ with odd-parity is depicted in
Fig.~\ref{fig2:6}, which is the reason why the probability
$|u_{k}(j)|^{2}$ is zero at $j=0$. If one regards the segment
between the two atoms as a finite chain with $N=2d$ sites, the wave
number in this segment takes $N$ discrete values, which give rise to
the discrete energy levels. Figure~\ref{fig2:6} shows that when the
transition energy $\Omega$ matches one of the discrete energy levels
in the segment sandwiched by the two atoms, \textit{bound states}
appear and a quantum super-cavity is formed. These bound states are
independent of the coupling strength $J$; however, a nonzero $J$ is
necessary.

\subsection{quantum super-cavity made of superconducting qubits}

Using superconducting charge qubits (one kind of ``artificial
atom'') as an example, we now focus on the question on trapping and
re-emitting photons in this unusual type of atomic resonator. It is
well known that the transition frequency $\Omega$ of superconducting
charge qubits can be controlled by both the voltage applied to the
gate and the external flux through the SQUID
loop~\cite{YouPT58,YouPRB68,liuPRA71}. Let us assume that a photon
with energy
\begin{equation}
E_{k}=E_{n}=\omega -2\xi \cos q_{n} 
\end{equation}
is initially in the $(-d)$th cavity. First, we tune the transition
frequency $\Omega$ outside the energy band and consider a large
detuning $(\Omega -E_{n})$. When the photon meets the first qubit,
it passes the first qubit and moves freely beyond the first qubit
due to the large detuning. After the photon is inside the spatial
range $\left[-d, \, d\right]$ between the two qubits, the transition
frequencies $\Omega_{1}=\Omega_{2}=\Omega$ are adjusted inside the
energy band, and satisfy Eq.~(\ref{rs-06}). Therefore the photon
would be totally trapped inside the super-cavity. We note that a
tunable super-cavity could also be
obtained by doping two $\Lambda$-type atoms inside the coupled-cavity array~%
\cite{gongzr}.

Based on the previous discussion in this paper, we can conclude the
following: (1) the single-photon can be trapped in the region $[-d, \, d]$
with a finite lifetime; (2) the single-photon can get out of the atomic
resonator when the transition frequencies $\Omega$ of these two atoms are
not equal to the incident energy of the single-photon. In the appendix we
will show that the existence of the photon bound states between the two
atomic mirrors is independent of the magnitude of the transition energy $%
\Omega$. We also conclude that a new cavity is formed by the two
atoms separately embedded in the two cavities of the coupled-cavity
array. Therefore, in analogy with superlattices in solid state, we
call this cavity a \textit{super-cavity} and the atoms act as
\textit{atomic mirrors}.

\section{\label{Sec:5}Long-wavelength effective theory}

In this section, we show that the real part of the momenta of the
quasi-bound levels in the low-energy region can be obtained by
expanding the sine and cosine functions in Eq.~(\ref{rs-05}) as
$\sin q_{n}\approx q_{n}$ and $\cos q_{n}\approx 1-q_{n}^{2}/2$.
Low-energy photons propagating along the resonator waveguide have
long wavelengths. Under the long-wavelength approximation, a
quadratic spectrum
\begin{equation}
E_{k}^{L}=\omega _{\xi }+\xi k^{2}  \label{loen-01}
\end{equation}%
is found by expanding the cosine function around zero in
Eq.~(\ref{eq:4}), where the superscript $L$ in $E_{k}^{L}$ refers to
the long-wavelength or lower energy regime and $\omega _{\xi
}=\omega -2\xi $. By introducing the field operator
\begin{equation}
\varphi \left( x\right) \equiv \int_{-\infty }^{\infty }dk\exp \left(
ikx\right) a_{k}
\end{equation}%
with the commutation relation
\begin{equation}
\left[ \varphi \left( x\right) ,\varphi ^{\dag }\left( x^{\prime }\right) %
\right] =\delta \left( x-x^{\prime }\right) ,
\end{equation}%
the Hamiltonian of the system in real space becomes
\begin{eqnarray}
H &=&\int_{-\infty }^{\infty }dx\varphi ^{\dag }\left( \omega _{\xi }-\xi
\partial _{x}^{2}\right) \varphi +\sum_{l}\left\{ \Omega \left\vert
e\right\rangle _{l}\left\langle e\right\vert \right.   \label{loen-02} \\
&&\left. +J\int_{-\infty }^{\infty }dx\delta \left[ x+\left( -1\right) ^{l}d%
\right] \left( \varphi ^{\dag }S_{l}^{-}+h.c.\right) \right\} ,  \notag
\end{eqnarray}%
where $S_{l}^{-}=\left\vert g\right\rangle _{l}\left\langle e\right\vert $
is the spin lowering operator of the $l$th atom. Since the total number of
excitations is conserved, we consider the storage of a single photon in the
region separated a distance $2d$ by two $\delta $-potentials. In the
coordinate representation, the stationary state of the system
\begin{eqnarray}
\left\vert E_{k}^{L}\right\rangle  &=&\int_{-\infty }^{\infty
}dx\;u_{k}\!\left( x\right) \varphi ^{\dag }\!\left( x\right) \left\vert
0gg\right\rangle   \label{loen-03} \\
&&+u_{k1e}^{\mathrm{long}}\left\vert 0eg\right\rangle +u_{k2e}^{\mathrm{long}%
}\left\vert 0ge\right\rangle   \notag
\end{eqnarray}%
is the superposition of a single photon (first term) and the
single-excited states of the two atoms (second and third terms). The
effective equation for the photon
\begin{eqnarray}
&&JG_{k}\left[ \delta \left( x-d\right) u_{k}\left( d\right) +\delta
\left( x+d\right) u_{k}\left( -d\right) \right]\label{loen-04} \\
&&=\left(\xi \partial _{x}^{2}+E_{k}^{L}-\omega _{\xi }\right)
u_{k}\left( x\right) \notag
\end{eqnarray}
is achieved from the eigenvalue equation $H\left\vert
E_{k}^{L}\right\rangle =E_{k}^{L}\left\vert E_{k}^{L}\right\rangle
$. Two $\delta $-potentials appear in Eq.~(\ref{loen-04}) along the
direction of the photon propagation, one is located at $x=-d$ and
the other is located at $x=d$. The height of the potential is
dependent on the energy carried by a single-photon. These two atoms
divide the region of photon propagation into three zones: (I)
$x<-d$; (II) $-d<x<d$; (III) $x>d$. The effective Hamiltonian
\begin{equation} H_{eff}=\omega _{\xi }-\xi \partial _{x}^{2},
\end{equation}
is valid in all three zones, and corresponds to free-particle
Schr\"{o}dinger equations, except for the replacement of $E_{k}^{L}$
by $(E_{k}^{L}-\omega _{\xi })$ in
$H|E_{k}^{L}\rangle=E_{k}^{L}|E_{k}^{L}\rangle$.

\begin{figure}[tbp]
\includegraphics[width=5 cm]{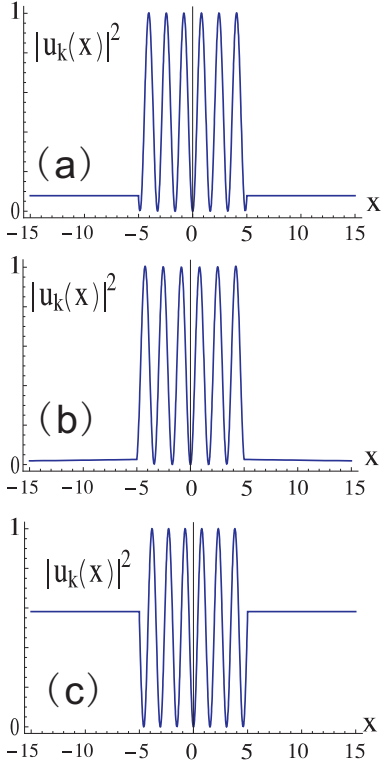}
\caption{(Color online). The probability $|u_{k}(x)|$ for finding
the photon in space using the long-wavelength effective theory. The
distance between qubits is $2d=10$. Other parameters are in units of
$J$. $\xi =0.1$, $n=3$, $\omega =5$ (a) $\Omega =3$, (b) $\Omega
=4$, (c) $\Omega =7$. For a given $n$, $q_{n}=\frac{n\pi}{d}$. These
were the inputs to equations~(\ref{loen-08}) and~(\ref{loen-08a}),
which provide $Q_{n}^{L}$ and $k$. With this $k$,
Eqs.~(\ref{loen-05}--\ref{loen-07}) are used to obtain the
$u_{k}(x)$'s shown in the Figures.} \label{fig2:7}
\end{figure}
We now concentrate on the case $E_{k}^{L}>\omega _{\xi }$. From the
standard boundary conditions that a wavefunction is always
continuous and its derivative is continuous except at points where
the potential is infinite, we can derive the continuity equations
for wave function $u_{k}\left( x\right)$ in different zones, and the
discontinuity of its derivatives (slopes) at the points $x=\pm d$.
According to the symmetry of the system, we assume that
Eq.~(\ref{loen-04}) has the following solution
\begin{equation}
u_{k}\left( x\right) =\left\{
\begin{array}{c}
S_{1}e^{-ikx}\text{, \ \ \ \ \ \ \ \ \ \ \ \ }x<-d, \\
e^{ikx}+B_{L}e^{-ikx}\text{, }-d<x<d, \\
S_{2}e^{ikx}\text{, \ \ \ \ \ \ \ \ \ \ \ \ \ }x>d.%
\end{array}%
\right.  \label{loen-05}
\end{equation}%
Using the same approach described in the section IV, we obtain the
coefficients
\begin{subequations}
\label{loen-06}
\begin{align}
S_{1}& =\frac{k\xi e^{-i2kd}+JG_{k}^{L}\sin \left( 2kd\right) }{2k\xi
+iJG_{k}^{L}}\text{,} \\
S_{2}& =\frac{k\xi }{2k\xi +iJG_{k}^{L}}\text{,} \\
B_{L}& =\frac{JG_{k}^{L}e^{i2kd}}{2ik\xi -JG_{k}^{L}}\text{,}
\end{align}%
with $G_{k}^{L}=J/\left( E_{k}^{L}-\Omega \right)$ and the condition for the
existence of the resonant states
\end{subequations}
\begin{equation}
e^{i2kd}=\pm \frac{2i\xi }{J^{2}}k\left( E_{k}^{L}-\Omega \right) \mp 1\text{%
.}  \label{loen-07}
\end{equation}
Here the wave number $k$ is complex. Under the condition $ Q=2\xi
^{2}/\left(J^{2}d^{3}\right) \ll 1$, the wave number
\begin{eqnarray}
k &=&q_{n}-\frac{1}{2}Q_{n}^{L}+d(Q_{n}^{L})^{2}  \notag \\
&&+i\frac{\lambda }{2d}Q_{n}^{L}\left( \delta _{\xi }+3\xi
q_{n}^{2}\right)+\mathcal{O}\left( Q^{3} \right) \label{loen-08}
\end{eqnarray}
is approximately obtained, up to second order in the parameter $Q$,
for those states with odd parity, and
\begin{equation}
Q_{n}^{L}=\frac{\lambda q_{n}}{d} \left( \omega -2\xi -\Omega +\xi
q_{n}^{2}\right) \label{loen-08a}\text{.}
\end{equation}
The super-index $L$ in Eqs.~(\ref{loen-08}--\ref{loen-08a}) refers
to the low-energy regime (long-wavelength approximation) studied in
this section. The real part, $\text{Re}(k)$, and the imaginary part,
$\text{Im}(k)$ in Eq.~(\ref{loen-08}) of a quasi-bound state provide
the energy and the lifetime of this state via the dispersion
relation in Eq.~(\ref{loen-01}). It is clear that the $\text{Re}(k)$
can be obtained by expanding the sine and cosine functions around
zero. Obviously, when $k=q_{n}$ and $\Omega =\omega _{\xi }+\xi
k^{2}$, a perfect cavity is formed. In this case, the coefficients $
S_{1}$ and $S_{2}$ are zero. In Fig.~\ref{fig2:7}, the probability
for finding a photon in space is shown. As the transition energy
$\Omega$ varies, the effective potential induced by the two qubits
changes from barriers to wells. It can be found that, as the depth
or height of the delta potential becomes larger, the leakage of the
super-cavity becomes smaller, which offers a way to control the
leakage of the super-cavity by adjusting the energy level spacing of
the two qubits. Therefore single-photons can be trapped.

\section{\label{Sec:6}short-wavelength effective theory}

In the higher-energy regime, the short-wavelength approximation
leads to a linear spectrum $E_{k}=\omega _{\pi }+2\xi \left\vert
k\right\vert $, with $\omega _{\pi }=\omega -\pi \xi $. Introducing
the left (right) bosonic field operator $\varphi _{L}^{\dag }\left(
x\right) $ ($\varphi _{R}^{\dag }\left( x\right) $), which creates a
left-moving (right-moving) particle at $x$, the tight-binding
Hamiltonian now becomes
\begin{eqnarray}
H_{c} &=&\omega _{\pi }\sum_{\alpha =R,L}\int_{-\infty }^{\infty }dx\varphi
_{\alpha }^{\dag }\left( x\right) \varphi _{\alpha }\left( x\right)
\label{hien-01} \\
&+&2i\xi \int_{-\infty }^{\infty }dx\left[ \varphi _{R}^{\dag }\left(
x\right) \partial _{x}\varphi _{R}\left( x\right) -\varphi _{L}^{\dag
}\left( x\right) \partial _{x}\varphi _{L}\left( x\right) \right] \text{.}
\notag
\end{eqnarray}
The left-moving and right-moving fields interact with these atoms
respectively, therefore the interaction Hamiltonian becomes
\begin{eqnarray}
H_{I} &=&J\sum_{\alpha l}\int_{-\infty }^{\infty }dx\delta \left(
x+(-)^{l}d\right) \left[ \varphi _{\alpha }^{\dag }\left( x\right)
S_{l}^{-}+h.c.\right]  \notag  \label{hien-02} \\
&&+\Omega \sum_{l}\left\vert e\right\rangle _{l}\left\langle e\right\vert.
\end{eqnarray}
Although the Hamiltonian in the short-wavelength regime (linear
dispersion regime) is significantly different from previous ones,
the number of total excitations is still a conserved quantity. The
stationary state for $H=H_{c}+H_{I}$ with one particle excitation
takes the form
\begin{eqnarray}
\left\vert E_{k}^{S}\right\rangle &=&\sum_{\alpha }\int
dx\;u_{k\alpha}\!\left(x\right) \varphi _{\alpha }^{\dag}\!\left(
x\right) \left\vert
0gg\right\rangle  \label{hien-03} \\
&&+u_{1e}\left\vert 0eg\right\rangle +u_{2e}\left\vert 0ge\right\rangle,
\notag
\end{eqnarray}
where the first number $0$ in Dirac bracket represents the vacuum
state of the cavity fields. Hereafter, the sub-index ``S'' in will
refer to the short-wavelength approximation regime. $u_{kR}\left(
x\right)$ and $u_{kL}\left( x\right)$ represent the probability
amplitudes for finding the photon along the right-moving and
left-moving direction at position $x$. Moreover, $u_{je}$ (with
$j=1,\,2$) are the probability amplitudes for one qubit in the
excited state and the other one in the ground state. From the
Schr\"{o}dinger equation, we obtain the relation between the
left-moving amplitude and the atomic amplitude in the excited state
\begin{equation}
\left( E_{k}^{S}-\omega _{\pi }-2i\xi \partial _{x}\right)
u_{kL}=J\sum_{j}\delta \left[ x+\left( -\right) ^{j}d\right]
u_{je}\text{.} \label{hien-04a}
\end{equation}
The relation between the right-moving amplitude and the atomic
amplitude is
\begin{equation}
\left( E_{k}^{S}-\omega _{\pi }+2i\xi \partial _{x}\right)
u_{kR}=J\sum_{j}\delta \left[ x+\left( -\right) ^{j}d\right]
u_{je}\text{.} \label{hien-04b}
\end{equation}
We can also find that the atomic amplitude $u_{je}$, the right-going
amplitude $u_{kR}$, and left-moving amplitudes $u_{kL}$ satisfy the
relation
\begin{equation}
u_{je}=G_{k}^{S}\int_{-\infty }^{\infty }dx\delta \left[ x+\left(
-\right) ^{j}d\right] \left( u_{kR}+u_{kL}\right),  \label{hien-04c}
\end{equation}
with the Green function $G_{k}^{S}=J/\left( E_{k}^{S}-\Omega
\right)$. After eliminating the variables $u_{1e}$ and $u_{2e}$,
both the left-moving eigenfunction
\begin{eqnarray}
&&\left( E_{k}^{S}-\omega _{\pi }-i2\xi \partial _{x}\right)
u_{kL}\left(
x\right)  \label{hien-05} \\
&=&JG_{k}^{S}\delta \left( x-d\right) \int_{-\infty }^{\infty
}dx^{\prime }\delta \left( x^{\prime }-d\right) \left[ u_{kR}\left(
x^{\prime }\right)
+u_{kL}\left( x^{\prime }\right) \right]  \notag \\
&&+JG_{k}^{S}\delta \left( x+d\right) \int_{-\infty }^{\infty
}dx^{\prime }\delta \left( x^{\prime }+d\right) \left[ u_{kR}\left(
x^{\prime }\right) +u_{kL}\left( x^{\prime }\right) \right]  \notag
\end{eqnarray}
and right-moving eigenfunction
\begin{eqnarray}
&&\left( E_{k}^{S}-\omega _{\pi }+i2\xi \partial _{x}\right)
u_{kR}\left(
x\right)  \label{hien-06} \\
&=&JG_{k}^{S}\delta \left( x-d\right) \int_{-\infty }^{\infty
}dx^{\prime }\delta \left( x^{\prime }-d\right) \left[ u_{kR}\left(
x^{\prime }\right)
+u_{kL}\left( x^{\prime }\right) \right]  \notag \\
&&+JG_{k}^{S}\delta \left( x+d\right) \int_{-\infty }^{\infty
}dx^{\prime }\delta \left( x^{\prime }+d\right) \left[ u_{kR}\left(
x^{\prime }\right) +u_{kL}\left( x^{\prime }\right) \right]  \notag
\end{eqnarray}
are subjected to a delta potential with singularities at $x=\pm d$.

\begin{figure}[tbp]
\includegraphics[width=8 cm]{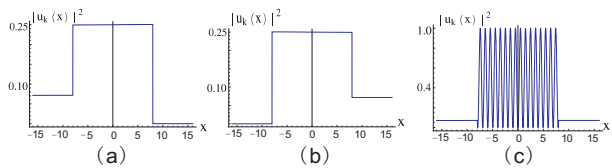}
\caption{(Color online). The norm square of the left-moving
wavefunction $|u_{kL}(j)|^{2}$ (a), the right-moving wavefunction
$|u_{kR}(j)|^{2}$ (b), and probability $|u_{k}(j)|^{2}\equiv
|u_{kL}(j)+u_{kR}(j)|^{2}$ (c) for finding the photon in space. The
parameters are set as follows: $d=8$, $\xi =0.1$, $n=1$, $\omega
=5$, $\Omega =2$. Parameters are in units of $J$. For a given $n$,
$q_{n}=\frac{n\pi}{d}$. These were the inputs to
equations~(\ref{hien-14}) and~(\ref{hien-14a}), which provide
$Q_{n}^{H}$ and $k$. With this $k$,
Eqs.~(\ref{hien-07}--\ref{hien-11}) are used to obtain the
$u_{k}(x)$'s shown in the Figs.(a)-(c)} \label{fig2:9}
\end{figure}

In the region $x\neq \pm d$, the potential is zero, and the
solutions of Eqs.~(\ref{hien-05}) and~(\ref{hien-06}) are plane
waves with left-moving and right-moving wave-vector number
$k=E_{k}^{S}/v_{g}$. Therefore, we can assume the right-moving
\begin{equation}
u_{kR}\left( x\right) =\left\{
\begin{array}{c}
0\text{, \ \ \ \ \ \ \ }x<-d, \\
e^{ikx}\text{, }-d<x<d, \\
t_{R}\;e^{ikx}\text{, \ \ \ \ \ \ }x>d,%
\end{array}%
\right.  \label{hien-07}
\end{equation}%
and the left-moving wave-function
\begin{equation}
u_{kL}\left( x\right) =\left\{
\begin{array}{c}
t_{L}\;e^{-ikx}\text{, \ \ \ \ }x<-d, \\
r_{L}\;e^{-ikx}\text{, }-d<x<d, \\
0\text{, \ \ \ \ \ \ \ \ \ }x>d,%
\end{array}%
\right.  \label{hien-08}
\end{equation}%
which allow the existence of quasi-bound states in this system. The
magnitude of $r_{L}$
\begin{equation}
r_{L}=\frac{JG_{k}^{S}}{i2\xi -JG_{k}^{S}}e^{i2kd}=\frac{i2\xi -JG_{k}^{S}}{%
JG_{k}^{S}}e^{-i2kd}  \label{hien-09}
\end{equation}%
and the relations
\begin{subequations}
\label{hien-10}
\begin{eqnarray}
t_{R} &=&r_{L}e^{-i2kd}+1 \\
t_{L} &=&r_{L}+e^{-i2kd}
\end{eqnarray}
of the amplitudes $t_{R}$, $t_{L}$ and $r_{L}$ can be obtained by
integrating Eqs.~(\ref{hien-05}) and~(\ref{hien-06}) in the
neighborhood of $x=\pm d$. For the appearance of quasi-bound states
in the spatial range sandwiched by two atoms, Eq.~(\ref{hien-09})
leads to the condition
\end{subequations}
\begin{equation}
e^{2ikd}=\pm \frac{2i\xi }{JG_{k}^{S}}\mp 1  \label{hien-11}
\end{equation}%
with the complex wave number $k$. Here, the lower sign corresponds
to the odd-parity, and the upper sign corresponds to the
even-parity. Obviously, when the transition energies $\Omega$ of the
two atoms are
\begin{equation}
\Omega =\omega _{\pi }+2\xi \left\vert \frac{n\pi}{d}\right\vert,
\label{hien-12}
\end{equation}
the bound states have odd parity. However the even parity corresponds to the
transition energy
\begin{equation}
\Omega =\omega _{\pi }+2\xi \left\vert \frac{\pi }{d}\left( n+\frac{1}{2}%
\right) \right\vert \text{.}  \label{hien-13}
\end{equation}%
Except the situation discussed above, Eq.~(\ref{hien-11}) does not
have an exact solution. We now seek the values of $k$ for which
Eq.~(\ref{hien-11}) can be approximately solved. Here we only
consider the energy levels with odd parity. A similar calculation
provides results for even parity. Using the approach described above
with the parameter $P=4\xi^{2}/(dJ^{2})$ and $\lambda=2\xi/J^{2}$,
Eq.~(\ref{hien-11}) with the lower sign, yields the wave number
\begin{equation}
k\approx q_{n}-\frac{J}{2d}Q_{n}^{S}+d(Q_{n}^{S})^{2}+i\xi
\frac{\lambda }{d}Q_{n}^{S}+\mathcal{O}\left(P^{3}\right)
\label{hien-14}
\end{equation}%
whose real part can be obtained from Eq.~(\ref{rs-05}) by expanding
the sine and cosine functions as $\sin q_{n}\approx 1$ and $\cos
q_{n}$ around $\pi /2$. Here,
\begin{equation}
Q_{n}^{S}= \frac{\lambda}{d} \left( \delta _{\pi }+2\xi q_{n}\right)
\label{hien-14a}
\end{equation}
and
\begin{equation}
\delta _{\pi }=\omega -\pi \xi -\Omega.
\end{equation}
We plot the norm square of the left-going wave-function in
Fig.~\ref{fig2:9}(a), the right-going wave-function in
Fig.~\ref{fig2:9}(b), and the total wave-function $u_{k}(j)$ in
Fig.~\ref{fig2:9}(c), where $u_{k}(j)\equiv u_{kL}(j)+u_{kR}(j)$.

\section{\label{Sec:7}Conclusions}

We have studied the coherent control of single photon transfer in a
coupled resonator waveguide with two atoms. The coherent control can
be realized by adjusting the detuning between the single photon
frequency and the energy-level-spacings of the atoms. We have shown
that a super-cavity is formed in the coupled-cavity array due to the
strong coupling between the atoms and the corresponding cavities,
and the discrete values of the photon momenta are analytically
derived. Moreover, a perfect super-cavity appears when the
transition energies of the two atoms are equal to the energy of an
incident photon. We also find that besides the bound states formed
by two perfect atomic mirrors, there always exist other bound states
at the edge of the band. The real parts of the discrete momenta
obtained by the discrete approach unify those obtained by the
effective continuum theory in both the long-wavelength and
short-wavelength regions.

\section{Acknowledgments}

This work is supported in part by NSFC Grants No.~90203018,
No.~10474104, No.~60433050, No.~10775048 and No.~10704023, NFRPC
Grant No.~2006CB921205, No.~2007CB925204, and No.~2005CB724508. F.N.
acknowledges partial support from the National Security Agency,
Laboratory Physical Science, Army Research Office, National Science
Foundation Grant No. EIA-0130383, JSPSRFBR Contract No.~06-02-91200,
and Core-to-Core program supported by the Japan Society for
Promotion of Science (JSPS). One of the authors (L.Z.) acknowledges
useful discussions with S. Ashhab.

\appendix

\section{Photon Bound states between two atomic mirrors}

To find the wave-functions for the eigenvalue
equation~(\ref{rs-01}), one needs to write down the wave-functions
in different regions. Since exchanging of the two atoms does not
change the equations for the wave-functions of the photon
propagating along the CRW, here we only deal with odd-parity wave
functions, which have the sinh function in the center region and
exponential decay in the edge regions
\begin{equation}
\Psi _{-}\left( x\right) =\left\{
\begin{array}{c}
-A\exp{[\left( in\pi +\kappa \right) j]}\text{ \ \ \ \ \ \ \ \ \ \ \ }x<-d, \\
B\exp{(in\pi j)}\sinh \left( \kappa j\right) \text{ \ }-d<x<d, \\
A\exp{[\left( in\pi -\kappa \right) j]}\text{\ \ \ \ \ \ \ \ \ \ \ }x>d.%
\end{array}%
\right.  \label{bound-01}
\end{equation}%
From the continuity and discontinuity conditions at $x=d$,
\begin{align*}
u\left( d^{+}\right) & =u\left( d^{-}\right), \\
\left( \omega +JG_{\kappa }-E\right) u\left( d\right) & =\xi \left[
u\left( d+1\right) +u\left( d-1\right) \right],
\end{align*}%
we can easily obtain
\begin{equation}
\tanh \left( \kappa d\right) =\frac{\xi \exp{(-in\pi)}\sinh \kappa
}{E-\omega -JG+\xi \left( e^{in\pi -\kappa }+e^{-in\pi }\cosh \kappa
\right) } \label{bound-02}
\end{equation}%
with
\begin{equation*}
E_{\kappa }=\omega -\xi \left( e^{in\pi -\kappa }+e^{-in\pi +\kappa }\right)
\end{equation*}%
and
\begin{equation*}
G_{\kappa }=\frac{J}{E_{\kappa }-\Omega}.
\end{equation*}
In principle, $\kappa$ can be obtained by solving the implicit
transcendental equation (\ref{bound-02}). It is obvious that
$\kappa=0$ is one of solution of Eq.~(\ref{bound-02}). This
$\kappa=0$ solution makes sure that the odd-parity wave functions
exist and two bound states appear at the edges.


\begin{thebibliography}{99}
\bibitem{Harris} S. E.~Harris, Phys. Today \textbf{50} (7), 36 (1997).

\bibitem{Lukin1} M.~Fleischhauer and M.D.~Lukin, Phys. Rev. Lett. \textbf{\
84}, 5094 (2000); Phys. Rev. A \textbf{65}, 022314 (2002).

\bibitem{scp03} C. P. Sun, Y. Li, and X. F. Liu, Phys. Rev. Lett \textbf{91}%
, 147903 (2003).

\bibitem{phoech1} T. W. Mossberg, Opt. Lett. \textbf{7}, 77 (1982).

\bibitem{phoech2} H. Lin, T. Wang, and T. W.~Mossberg, Opt. Lett. \textbf{20}%
, 1658 (1995).

\bibitem{CRIB} B. Kraus, W. Tittel, N. Gisin, M. Nilsson, S. Kr\"{o}ll, and
J. I. Cirac, Phys. Rev. A \textbf{73}, 020302(R) (2006).

\bibitem{cavity1} K. J. Vahala, Nature \textbf{424}, 839 (2003).

\bibitem{cavity3} J. Bravo-Abad and M. Solja\v{c}i\'{c}, Nature Mater.
\textbf{6}, 799 (2007)

\bibitem{cavity4} Y. Tanaka, J. Upham, T. Nagashima, T. Sugiya, and T.
Asano, Nature Mater. \textbf{6}, 862 (2007).

\bibitem{cavity4a} M. Sandberg, C. M. Wilson, F. Persson, T. Bauch, G. Johansson,
V. Shumeiko, T. Duty, and P. Delsing, Appl. Phys. Lett. \textbf{92},
203501 (2008).

\bibitem{cavity4b} M. A. Castellanos-Beltran and K. W. Lehnert,
Appl. Phys. Lett. \textbf{91}, 083509 (2007).

\bibitem{cavity} J. Kim, O. Benson, H. Kan, and Y. Yamamoto, Nature \textbf{
397}, 500 (1998).

\bibitem{cavity2} B. Dayan, A. S. Parkins, T. Aoki, E. P. Ostby, K. J.
Vahala, and H. J. Kimble, Science \textbf{319}, 1062 (2008).

\bibitem{cavity5} F. Y. Hong and S. J. Xiong, Phys. Rev. A \textbf{78},
013812 (2008).

\bibitem{Lukin-np} D. E. Chang, A. S. S{\o }rensen, E. A. Demler, and M. D.
Lukin, Nat. Phys. \textbf{3}, 807 (2007).

\bibitem{fanpaper} J. T. Shen and S. Fan, Phys. Rev. Lett. \textbf{95},
213001 (2005); \textit{ibid.} \textbf{98}, 153003 (2007); Opt. Lett. \textbf{%
30}, 2001 (2005).

\bibitem{ZGLSN} L. Zhou, Z. R. Gong, Y. X. Liu, C. P. Sun, and F. Nori,
Phys. Rev. Lett. \textbf{101}, 100501 (2008).

\bibitem{cavity6} H. Dong, Z. R. Gong, H. Ian, L. Zhou, and C. P. Sun,
arXiv:0805.3085.

\bibitem{Sun1} C. P. Sun, L. F. Wei, Y. X. Liu, and F. Nori, Phys. Rev. A
\textbf{73}, 022318 (2006);

\bibitem{Mari} M. Mariantoni, F. Deppe, A. Marx, R. Gross, F. K. Wilhelm,
and E. Solano, Phys. Rev. B \textbf{78}, 104508 (2008).

\bibitem{greentree} A.~Greentree, C.~Tahan, J.H. Cole, and L.C.L.
Hollenberg, Nat. Phys.~\textbf{2}, 856 (2006).

\bibitem{AngelA76} D. G. Angelakis, M. F. Santos, and S. Bose,
Phys. Rev. A \textbf{76}, 031805 (2007).

\bibitem{zhjsun76} L. Zhou, J. Lu, and C. P. Sun, Phys. Rev. A \textbf{76}, 012313 (2007).

\bibitem{hu76PRA} F.M. Hu, L. Zhou, T. Shi, and C. P. Sun, Phys. Rev. A
\textbf{76}, 013819 (2007).

\bibitem{Plenio} M.J.~Hartmann, F. G. S. L. Brand\~{a}o, and M.B. Plenio,
 Nat. Phys.~\textbf{2}, 849 (2006).

\bibitem{Sergey} A.L.~Rakhmanov, A. M. Zagoskin, S. Savel'ev, and F. Nori,
Phys. Rev. B \textbf{77}, 144507 (2008).

\bibitem{FNRMP} K. Yu. Bliokh, Yu. P. Bliokh, V. Freilikher, S. Savel'ev, and F.
Nori, arXiv:0708.2653. Rev. Mod. Phys., in press.

\bibitem{fcitaly} F. Ciccarello, G. M. Palma, M. Zarcone, Y. Omar, V. R.
Vieira, J. Phys. A: Math. Theor. \textbf{40}, 7993 (2007); Las.
Phys. \textbf{17}, 889 (2007); New J. Phys. \textbf{8}, 214 (2006)

\bibitem{quaspr} F. M. Dittes, Phys. Rep. \textbf{339}, 215 (2000).

\bibitem{arkaB72} Y. S. Joe, A. M. Satanin, and G. Klimeck, Phys. Rev. B \textbf{72},
115310 (2005).

\bibitem{scat-pole} N. Hatano, K. Sasada, H. Nakamura and T. Petrosky,
Prog. Theor. Phys. \textbf{119}, 187 (2008); e-print
arXiv:0705.1388.

\bibitem{YouPT58} J. Q. You and F. Nori, Phys. Today \textbf{58} (11), 42
(2005).

\bibitem{YouPRB68} J. Q. You and F. Nori, Phys. Rev. B \textbf{68}, 064509
(2003).

\bibitem{liuPRA71} Y. X. Liu, L. F. Wei, and F. Nori, Phys. Rev. A \textbf{\
71}, 063820 (2005).

\bibitem{gongzr} Z. R. Gong, H. Ian, L. Zhou, and C. P. Sun, arXiv:0805.3042.
\end{thebibliography}
\end{document}